%% file: pr363.tex
\documentclass[12pt]{article}
\usepackage{epsfig}
\usepackage{a4p}
\usepackage{cite}
\usepackage{mcite}

\renewcommand{\arraystretch}{1.1}

\begin{document}
\def\be{\begin{equation} }
\def\ee{\end{equation} }
\def\ba{\begin{eqnarray} }
\def\ea{\end{eqnarray} }
\def\ban{\begin{eqnarray*} }
\def\ean{\end{eqnarray*} }
\def\epem{\mbox{e}^+\mbox{e}^-}
\def\eegg{\epem\to\gamma\gamma}
\def\eeggg{\epem\to\gamma\gamma(\gamma)}
\def\ggg{\gamma\gamma(\gamma)}
\def\ct{\cos{\theta}}
\def\cte{\cos{\theta^{\ast}}}
\def\pl{p_{\rm l}}
\def\pt{p_{\rm t}}
\def\g{\gamma}
\def\B{{\mathcal{B}}}
\def\O{{\mathcal{O}}}
\def\R{{\mathcal{R}}}
\def\E{{\mathcal{E}}}
\def\Lpm{\Lambda_{\pm}}
\def\xmc{\left(\frac{d\sigma}{d\ct}\right)_{\rm MC}}
\def\xb{\left(\frac{d\sigma}{d\ct}\right)_{\rm Born}}
\def\xl{\left(\frac{d\sigma}{d\ct}\right)_{\Lambda_{\pm}}}
\def\xq{\left(\frac{d\sigma}{d\ct}\right)_{\rm \Lambda '}}
\def\xe{\left(\frac{d\sigma}{d\ct}\right)_{\rm e^{\ast}}}
\def\xg{\left(\frac{d\sigma}{d\ct}\right)_{M_s}}
\def\xsn{\frac{d\sigma}{d\ct}}
\def\ca{$I$}
\def\cb{$I\!I$}
\def\cc{$I\!I\!I$}
\def\cd{$I\!V$}
\def\acol{\xi_{\rm acol}}
\def\acop{\xi_{\rm aplan}}
\def\athet{\xi_{\theta}}
\def\mestar{M_{\rm e^\ast}}
\def\ng{N_{\gamma}}
\def\Es{E_S}
\def\pt{p_{\,\rm t}}
\def\pl{p_{\,\rm l}}

\begin{titlepage}
\begin{center}{\large   EUROPEAN ORGANIZATION FOR NUCLEAR RESEARCH
 }\end{center}\bigskip
\begin{flushright}
CERN-EP-2002-060 \\
17 July 2002\\
\end{flushright}
\bigskip\bigskip\bigskip\bigskip\bigskip
\begin{center}
 {\LARGE\bf \boldmath
Multi-Photon Production in $\epem$ Collisions\\[0.5ex]
at $\sqrt{s} = 181 - 209 \mbox{ GeV}$ }
\end{center}
\bigskip\bigskip

\begin{center}{\LARGE The OPAL Collaboration}
\end{center}\bigskip\bigskip
\bigskip\begin{center}{\large  Abstract}\end{center}


The process $\eeggg$ is studied using data collected by the OPAL 
detector at LEP between the years 1997 and 2000. 
The data set corresponds to an integrated luminosity 
of 672.3~pb$^{-1}$ at centre-of-mass energies lying 
between 181 GeV and 209 GeV.
Total and differential cross-sections are determined and found to be in
good agreement with the predictions of QED. Fits to the observed 
angular distributions are used to set limits 
on parameters from several models of physics beyond the Standard Model 
such as cut-off parameters, contact interactions of the type
$\epem\gamma\gamma$, gravity in extra spatial dimensions and  
excited electrons. 
In events with three photons in the final state the mass spectrum 
of photon pairs is investigated. No narrow resonance $X\to\gamma\gamma$ is 
found and limits are placed on the product of the 
$\rm X \gamma$ production cross-section and branching ratio.

 \bigskip
\begin{center}

 \bigskip
 \bigskip

 \end{center}
\begin{center}
{\large ( To be submitted to the Eur. Phys. J. C)}
 \end{center}
\end{titlepage}

\input collaboration.tex

\section{Introduction}

This paper presents a study of the process $\eeggg$, where
the brackets indicate a possible third particle that might escape 
along the beam direction.
Data with a total integrated luminosity of 672.3 pb$^{-1}$
collected in the years 1997 to 2000 with the OPAL detector at LEP
at centre-of-mass energies between 181 GeV and 209 GeV are used.
This process is one of the few reactions which are dominated by QED 
even at these energies. The high statistics sample considered in this 
paper allows precision tests of QED and searches for new particles such 
as excited electrons and photonically decaying resonances.

This process has been studied previously at LEP at lower energies
\cite{ref:ich_172,ref:ich_183,ref:ich_189} and by other 
experi\-ments
\cite{ref:ggaleph_2000,ref:ggaleph_183,ref:ggdelphi_183,ref:ggdelphi_1999,ref:ggl3_2000,ref:ggl3_189,ref:ggl3_172,ref:ggl3_130}.
The increased data sample since our last publication leads
to a much better understanding of the detector and the 
systematic errors. In addition the selection efficiency has been
increased. Therefore this paper includes a re-analysis of previously
published data at 183 GeV and 
{\mbox{189 GeV \cite{ref:ich_183,ref:ich_189}} and supersedes those
results.
Total and differential cross-sections are determined and compared with
the predictions of QED.

There are several models which predict deviations from the 
QED cross-section introducing cut-off parameters \cite{ref:drell}, 
$\epem\gamma\gamma$ contact interactions \cite{ref:eboli}, the 
exchange of excited electrons \cite{ref:low,ref:estar} or the exchange 
of Kaluza-Klein gravitons in models with extra dimensions
\cite{ref:exdim_AD}. The measured differential cross-sections are 
used to place constraints on the parameters of these models.

The search for a resonance X which is produced along with a photon
and decays via $\rm X\to\gamma\gamma$ is especially interesting in the
light of models with fermiophobic Higgs bosons~\cite{ref:eboli_higgs} 
or hypercharge axions~\cite{ref:ram}. Because of the higher kinematic 
reach the three-photon final state is complementary to
$\rm ZX$ production. Events with
three observed photons are used to search for such a resonance.
Similar searches were performed previously at the Z
peak~\cite{ref:ggopal_Z_Hgg} and at lower LEP2 
energies~\cite{ref:ggdelphi_Hgg,ref:ich_172,ref:ich_183,ref:ich_189}.

In the next section a short description of the relevant detector
components is given. This is followed by an overview of the data and
Monte Carlo samples used in the analysis. Section~\ref{sec:theo} 
contains an introduction to 
the theoretical models. The event selection is described in 
Section~\ref{sec:select}, the systematic errors are summarised in 
Section~\ref{sec:syserr} and the results are given in 
Section~\ref{sec:results}.

\section{The OPAL detector}
\label{sec:det}

A detailed description of the OPAL detector can be found elsewhere
\cite{opal:det,opal:si} so only those components most relevant to 
this analysis are discussed here.

The most important subdetector for this analysis is the electromagnetic 
calorimeter (ECAL).
The barrel region\footnote{OPAL uses a right-handed 
coordinate system in which the $z$ axis is along the electron beam 
direction, $x$ points to the centre of LEP such that the $y$ axis points 
approximately upwards. The polar angle, $\theta$, is measured with 
respect to the $z$ axis and the azimuthal angle, $\phi$, with respect 
to the $x$ axis.} ($|\ct|<0.82$) consists of 9440 lead-glass blocks in a
quasi-pointing geo\-metry each with a cross-section of about 
$10 \times 10$ cm at an inner radius of 2.45 m. 
The two endcaps \mbox{($0.81<|\ct|<0.98$)} consist of 1132 blocks each,
aligned parallel to the beam axis. The inner faces 
are placed at a $z$ position of about $\pm$2.3 m.
For beam-energy photons, the spatial resolution is about 11 mm,
corresponding to an uncertainty of 0.2$^\circ$ in $\theta$, and the 
energy resolution is about 2\% in the barrel and  3\% -- 5\% in the 
endcaps, depending on the polar angle.

The ECAL surrounds the tracking chambers. The large volume jet chamber 
(CJ) has an outer diameter of 3.7 m and is about 4 m long. 
It is segmented into 24 sectors each containing 159 axial sense wires. 
The vertex chamber (CV) has an outer diameter of 47 cm and is 1 m long.
It is segmented into 36 sectors each containing 12 axial wires in the
inner region and
an outer region of 6 stereo wires. CV and CJ are inside a common 
pressure vessel and separated by a foil and a carbon fibre tube.
A silicon micro-vertex detector (SI) is 
located between CV and the beam pipe, at a radius of about 7 cm.
This subdetector consists of two layers of double sided 
($z$ and $\phi$ sensitive) ladders. The inner (outer) layer covers 
\mbox{$|\ct|<$ 0.93 (0.89)} and has a $\phi$ acceptance of 
97.8\% (99.6\%).

Outside the ECAL the instrumented  return yoke serves as a
hadronic calorimeter. The outermost subdetector is the
muon system which consists of up to 4 layers of drift chambers.
The luminosity is measured using small-angle Bhabha events collected in
the silicon-tungsten luminometer \cite{ref:lumi}.

\section{Data sample and Monte Carlo simulation} 
\label{sec:mc}

\begin{table}[bp]
\begin{center}
\begin{tabular}{c|c|r@{$-$}l|c|r@{$\pm$}c@{$\pm$}l|r}
\hline
year & $\sqrt{s}_{\rm nom}$ & 
\multicolumn{2}{c|}{$\sqrt{s}_{\rm range} \; [ \mbox{GeV} ]$} &
 $\sqrt{s} \; [ \mbox{GeV} ]$ &
\multicolumn{3}{c|}{luminosity $[ \mbox{pb}^{-1} ]$} & events \\
\hline\hline
  1997 & 183 & 180.8&184.0 & 182.68 &  55.57&0.15&0.19 & 538\\
  1998 & 189 & 188.6&189.0 & 188.63 & 181.07&0.16&0.36 & 1531\\
  1999 & 192 & 191.4&192.0 & 191.59 &  29.03&0.06&0.07 & 258\\
  1999 & 196 & 195.2&196.0 & 195.53 &  75.92&0.10&0.16 & 616\\
 99/00 & 200 & 199.4&200.2 & 199.52 &  78.20&0.11&0.17 & 554\\
 99/00 & 202 & 201.4&202.5 & 201.63 &  36.78&0.07&0.08 & 281\\
  2000 & 205 & 202.5&205.5 & 204.88 &  79.22&0.11&0.17 & 566\\
  2000 & 207 & 205.5&209.0 & 206.56 & 136.49&0.14&0.30 & 891\\  
\hline
\end{tabular}
\caption{Data used in this paper. The year of data taking, 
the nominal centre-of-mass energy, the energy range, the luminosity-weighted 
mean centre-of-mass energy, the integrated luminosity with
its statistical and systematic error and the number of selected events
are shown.}
\label{tab:sample}
\end{center}
\end{table}
Table \ref{tab:sample} lists the data samples analysed in this paper.
They are from the last four years of LEP running with a total
of 5235 selected events taken at centre-of-mass energies between 181~GeV
and 209~GeV. As an example, the last $\eeggg$ event, which was recorded 
three minutes before the final shut down of LEP, is shown in 
Figure \ref{fig:event}. Two photons, back-to-back in $\phi$, are observed
in the detector. One photon has converted between CV and CJ. 
From the acollinearity a third photon is deduced
to have escaped along the beam direction.

Various Monte Carlo samples are used to study the selection efficiency
and expected background contributions. The signal $\eeggg$ 
events are generated using the RADCOR \cite{ref:radcor} generator
which relies on a full ${\cal O}(\alpha^3)$ calculation taking the
electron mass into account. Events with four observed photons can be
simulated using FGAM \cite{ref:fgam} which contains a lowest-order 
relativistic calculation of four-photon production.
Bhabha events are simulated using BHWIDE \cite{ref:bhwide}
($\epem(\gamma)$) and TEEGG \cite{ref:teegg} ($\rm e\gamma (e)$). 
For $\epem\to\nu\bar{\nu}\gamma(\gamma)$ events KORALZ \cite{ref:koralz}
and NUNUGPV \cite{ref:nunugpv} are used.
Tau pairs are simulated using KORALZ \cite{ref:koralz}, and
PYTHIA \cite{ref:pythia6.2,ref:pythia6.3} is used for hadronic events. 
All samples are processed through the OPAL detector simulation program
\cite{ref:gopal} and reconstructed in the same way as the data.

\section[Cross-section for the process $\eeggg$]
        {Cross-section for the process \boldmath$\eeggg$}
\label{sec:theo}

\subsection{QED Born cross-section}

Up to the highest LEP energies the process $\eeggg$ can be described 
by QED. The lowest order cross-section in the relativistic limit is 
given by:
\be
\xb =  
\frac{2\pi \;\alpha^2}{s}\;\frac{1+\cos^2{\theta} }{1-\cos^2{\theta} } \; ,
\label{born}
\ee
where $\sqrt{s}$ is the centre-of-mass energy and $\alpha$ is the 
fine-structure constant at zero momentum transfer. 
The non-relativistic cross-section formula is given in \cite{ref:QED}.
Since the final-state particles are identical, the polar angle $\theta$ 
is defined such that $\ct>0$. At Born level the definition of $\theta$ 
is unambiguous since the two photons are back-to-back. The definition
at higher orders used in this paper is given in Section
\ref{sec:radcor}.

Weak interactions contribute only via loop diagrams.
At the W-pair threshold weak corrections of up to 1.2\% 
are expected for $\ct=0$~\cite{ref:ggweak,ref:ggweak2}. 
At the energies considered in this analysis the 
corrections are smaller, e.g. for a centre-of-mass energy of 200 GeV 
they are less than 0.2\% for all angles, and will be neglected.

\subsection{Radiative corrections}
\label{sec:radcor}

In the presence of higher order effects the measured 
angular distribution depends not only on one 
angle $\theta$ but on the angles of all produced photons. 
To enable a comparison of the measurement with 
Equation \ref{born} an event angle must be defined to substitute 
$\theta$.  Various definitions of the event angle are possible,
each leading to a different measured angular distribution.
The ratio of this physical distribution to the distribution from
the Born-level prediction is called a radiative correction $\R$. 
The event angle $\theta^\ast$ used in this analysis is defined by:
\be
\cos{\theta^{\ast}} = \left|\sin{\frac{\theta_1 - \theta_2}{2}}\right|
     \; {\Bigg /} \; \left( {\sin{\frac{\theta_1 + \theta_2}{2}}}\right)
     \; ,
\label{ctstar} 
\ee
where $\theta_1$ and $\theta_2$ are the angles of the two 
highest-energy photons. At Born level $\ct = \cte$. 
The angle $\theta^\ast$ is equivalent to the angle in 
the centre-of-mass system of the two highest-energy photons unless a 
third photon is produced away from the beam direction. 
With this definition, deviations of $\R$ from unity are  
relatively small and uniform as determined from an $\O(\alpha^3)$ 
Monte Carlo~\cite{ref:radcor} without detector simulation. Using high
statistics samples of 10$^9$ events, generated at centre-of-mass 
energies of 189 GeV, 200 GeV and 206 GeV,  the correction $\R$ is 
found to depend only weakly on the centre-of-mass energy. Therefore the
$\sqrt{s}$ dependence is neglected. The average correction for
$\cte < 0.93$ is $\R = 1.0448$; the angular dependence can be found in
Table~\ref{tab:difxs}. For QED processes the effects due to the next
order can be assumed to be about 10\% of the corrections at this order.
Since no fourth order Monte Carlo generator is available and weak contributions 
are neglected, a systematic error of 0.01 is assumed for the radiative 
correction. The ratio $\R$ is used to correct the measured 
cross-sections presented in this paper to the Born level.

\subsection{Alternative models}

Various models predict deviations from the 
QED expectation\footnote{
All cross-sections given here are to lowest order.}.
The simplest ansatz is a short-range exponential deviation from the 
Coulomb field
parameterised by cut-off parameters $\Lpm$~\cite{ref:drell}. 
This leads to a differential cross-section of the form
\be
\xl   =  \xb \pm \frac{\alpha^2 \pi s}{\Lambda_\pm^4}(1+\cos^2{\theta}) \; .
\label{lambda}
\ee

New effects can also be introduced in effective Lagrangian theory
\cite{ref:eboli}. Here dimension-6 terms lead to anomalous 
$\rm ee\gamma$ couplings. The resulting deviations in the differential 
cross-section are similar in form to those given in 
Equation~\ref{lambda}, but with a slightly different definition of the
parameter: $\Lambda_6^4 = \frac{2}{\alpha}\Lambda_+^4$.
Dimension 7 and 8 Lagrangians introduce $\rm ee\gamma\gamma$ contact
interactions and result in an angle-independent term added to the Born
cross-section:
\be
\xq  =  \xb + \frac{s^2}{16}\frac{1}{\Lambda'{}^6} \; .
\ee
The associated parameters are given by 
$\Lambda_7 = \Lambda'$ and $\Lambda_8^4 = m_{\rm e} {\Lambda'}^3$ for
dimension~7 and dimension~8 couplings, respectively.
The subscript refers to the dimension of the Lagrangian.

Instead of an ordinary electron, an excited electron $\rm e^\ast$
with mass $\mestar$
could be exchanged in the $t$-channel \cite{ref:low,ref:estar}. 
In the most general case $\rm \rm e^\ast e \gamma$ couplings would lead
to a large anomalous magnetic moment of the electron 
\cite{ref:g2_brodsky,ref:g2_renard}. 
This effect can be avoided by a chiral magnetic coupling of the form
\begin{equation}
{\cal L}_{\rm e^\ast e \gamma} = 
\frac{1}{2\Lambda} \bar{e^\ast} \sigma^{\mu\nu}
\left[ g f \frac{\tau}{2}W_{\mu\nu} + g' f' \frac{Y}{2} B_{\mu\nu}
\right] e_L + \mbox{h.c.} \; ,
\end{equation}
where $\tau$ are the Pauli matrices and $Y$ is the hypercharge.
The parameters of the model are the compositeness scale $\Lambda$
and the weight factors $f$ and $f'$ associated to the gauge fields 
$W$ and $B$ with Standard Model couplings $g$ and $g'$.
For the process $\eeggg$, 
the following cross-section results~\cite{ref:vachon}: 

\begin{samepage}
\ba
\xe & = & \xb  \\
 & + & \frac{\alpha^2 \pi}{2}\frac{f_\gamma^4}{\Lambda^4}\mestar^2 \left[
\frac{p^4}{(p^2-\mestar^2)^2} + \frac{q^4}{(q^2-\mestar^2)^2} +
\frac{\frac{1}{2} s^2 \sin^2\theta}{(p^2-\mestar^2)(q^2-\mestar^2)} \right]
\; , \nonumber \ea
\end{samepage}
with $f_\gamma = -\frac{1}{2}(f+f')$, $p^2=-\frac{s}{2}(1-\ct)$ and 
$q^2=-\frac{s}{2}(1+\ct)$. 
Effects vanish in the case of $f = -f'$.

Theories of quantum gravity in extra spatial dimensions could solve the 
hierarchy problem because gravitons would be allowed to travel in 
more than 3+1 space-time dimensions \cite{ref:exdim}. 
While in these models the Planck mass $M_D$
in $D=n+4$ dimensions is chosen to be of electroweak scale the usual
Planck mass $M_{\rm Pl}$ in four dimensions would be
\be M_{\rm Pl}^2 = R^n M_D^{n+2} \; ,\ee
where $R$ is the compactification radius of the additional dimensions.
Since gravitons couple to the energy-momentum tensor, their
interaction with photons is as weak as with fermions. However, the huge
number of Kaluza-Klein excitation modes in the extra dimensions may 
give rise to
observable effects. These effects depend on the scale $M_s (\sim M_D)$ 
which may be as low as ${\cal O}(\rm TeV)$. Model dependencies
are absorbed in the parameter $\lambda$ which is expected to be 
of ${\cal O}(1)$. For this analysis it is assumed that $\lambda = \pm 1$. 
The expected differential cross-section is given by \cite{ref:exdim_AD}:
\begin{equation}
\xg = \xb - {\alpha s} \; \frac{\lambda}{M_s^4}\;(1+\cos^2{\theta})
    + \frac{s^3}{8 \pi} \;  \frac{\lambda^2}{M_s^8} \;(1-\cos^4{\theta})
    \; .
\end{equation}

\section{Event selection}
\label{sec:select}

Multi-photon events have a very clear experimental signature.
They have large energy deposits in the electromagnetic calorimeter and
small missing transverse momentum. They share these properties with
Bhabha events ($\epem\to\epem (\gamma)$), which represent the most 
important background. The first stage of the event selection 
is a preselection which requires events to have photon candidates and a
low multiplicity of tracks and clusters. Cuts are then applied to 
reject cosmic-ray background.
This is followed by a kinematic selection based on the signature of
the energy deposit in the ECAL. The last step is a neutral 
event selection designed to reject Bhabha events using information 
from the tracking chambers. After application of all selection criteria
the total background is reduced to an almost negligible level. 
\subsection{Preselection}

A photon candidate is defined here as an ECAL cluster with an energy 
of at least 1 GeV
and a polar angle satisfying $|\ct|<0.97$. The cluster must consist of at least
two ECAL blocks to ensure a good determination of the photon angle.

The preselection is very loose, requiring at least two photon candidates
and a total ECAL energy in the event of at least $0.1 \sqrt{s}$.
High multiplicity events are rejected by restricting the sum of the 
number of tracks and ECAL clusters to $\leq 17$.

The two highest-energy photons must be within $|\ct|<0.93$. This
avoids a region where the detector material is not sufficiently well
modelled to ensure a good description of the conversion probability and
the angular reconstruction. The search for a resonance X is based on 
three-photon events and the invariant mass of X is calculated
from the photon angles. Only events in which all three photons are in 
the region $|\ct|<0.93$ are used for this search.

\subsection{Cosmic-ray background rejection}

A cosmic-ray particle can create signals in the outer detectors
without producing a reconstructed track in the central
tracking chambers. These particles do not necessarily pass close to 
the beam axis. Since the hadronic and electromagnetic calorimeters
have different radii, the resulting hits in  the two detectors occur 
separated in azimuth. Background of this type is suppressed using 
information 
from the muon chambers and the hadronic calorimeter. Events are rejected 
if there are three or more track segments reconstructed in the muon
chambers. Events are also rejected if the highest-energy hadronic
cluster is separated from each photon candidate by at least 10$^\circ$ 
in $\phi$ and has an energy of more than 20\% of the summed energy 
of the photon candidates. In the case of one or two muon track segments 
the latter cut is tightened to 10\%.
Additionally, events are rejected if there is a track with a momentum 
of more than 10 GeV separated from each of the photon candidates by
more than $10^\circ$ in $\phi$.

\subsection{Kinematic event selection}

\begin{table}[bp]
\begin{center}
\renewcommand{\arraystretch}{1.5}
\begin{tabular}{|c|c|c|c|}
\hline
$\acol<10^\circ$ & \multicolumn{1}{c}{$\acol>10^\circ$} & \multicolumn{2}{c|}{} \\
\hline
 & \cb & \multicolumn{2}{r|}{$\ng = 2$} \\
\cline{2-4}
 & \cc & $\acop<0.1^\circ$ & $\ng$ \\
\cline{2-3}
 \raisebox{2.2ex}[-2.2ex]{\ca} &  & $\acop>0.1^\circ$ & $=3$ \\
\cline{3-4}
 & \raisebox{2.2ex}[-2.2ex]{\cd} & \multicolumn{2}{r|}{$\ng \geq 4$} \\
\hline
\end{tabular} 
\caption{Definition of classes \ca , \cb , \cc\ and \cd . All collinear
events are contained in \mbox{class \ca .} Other events are distributed
according to the number of photon candidates $\ng$ and the aplanarity
$\acop$.}
\label{tab:class}
\end{center}
\end{table}

Events are selected if they have large total energy and small missing
transverse momentum. To improve the resolution on these quantities the 
information on 
the angle of the cluster is used instead of the energy information where
possible. To facilitate this, the events are divided into four 
classes (\ca , \cb , \cc\ and \cd ) 
according to the number of photon candidates $\ng$, the 
acollinearity $\acol = 180^\circ - \alpha_{12}$ and the aplanarity 
$\acop = 360^\circ - (\alpha_{12} + \alpha_{13} + \alpha_{23})$,
where $\alpha_{ij}$ is the angle between photon candidates $i$ and $j$. 
Here and in the following sections the photon candidates are ordered 
by energy. The class definitions are shown in Table \ref{tab:class}.

Cuts are applied on the energy sum as well as on the missing transverse and 
longitudinal momenta. The energy sum $\Es$ is defined as the sum 
of the photon candidate energies $E_i$ plus the missing longitudinal momentum 
which could originate from a photon escaping along the beam direction.
The detailed definitions of these quantities depend on the event class. 
However the cuts summarised in Table \ref{tab:kincuts} namely
that the energy sum must be at least 60\% of the centre-of-mass energy
and the missing transverse momentum must be less than 
10\% of the centre-of-mass energy are equivalent.
Since the event angle is defined using the two highest-energy photons,
events are rejected if one of these escapes detection, that is if the
missing longitudinal momentum is larger than the energy of 
the second highest-energy photon candidate.
This cut on the missing longitudinal momentum is not designed to
reject background but ensures a good signal definition since it 
prevents ambiguous events from entering the signal sample.
Distributions of the relevant quantities are shown in Figure 
\ref{fig:kin}.
  
Class \ca\ contains collinear events which make up about 90\% of the
total sample. For these events it is assumed that missing transverse and longitudinal
momenta are negligible, so no cuts on these quantities are applied. 
The energy sum is simply taken to be $E^{I}_S = E_1 + E_2$.  A
class \cb\ event has exactly two observed photon candidates; since
the event is acollinear a third photon is assumed along the beam 
direction. Quantities $\B$ and $E_{\rm lost}$, equivalent to the 
missing transverse and longitudinal momenta, are calculated assuming
three-body kinematics and are given by:
\be \B = \sqrt{s}/2 \cdot (\sin{\theta_1}+\sin{\theta_2})
\left| \cos{\left[(\phi_1-\phi_2)/2\right]}\right| \; , \label{eq:B}\ee
\be E_{\rm lost} =  \sqrt{s} / \left[ 1 + (\sin{\theta_1} +
\sin{\theta_2}) /|\sin{(\theta_1 + \theta_2)}|\right] \; .\label{eq:lost}\ee

Consequently $E^{I\!I}_S = E_1 + E_2 + E_{\rm lost}$.
For classes \cc\ and \cd\ the missing transverse and longitudinal momenta, 
$\pt$ and $\pl$, are calculated in the usual way from cluster energies 
and angles and $E^{I\!I\!I}_S = \sum_i E_i + \pl$.

\begin{table}[htp]
\begin{center}
\begin{tabular}{l|ccc|c}
\hline
event class & \ca & \cb & \cc , \cd & cut \\\hline\hline
 & \rule{3em}{0em} & \rule{3em}{0em} & \rule{3em}{0em} & \\[-2ex]
energy sum  
     & $E^{I}_S$ & $E^{I\!I}_S$ & $E^{I\!I\!I}_S$ & $>0.6\sqrt{s}$ \\
transverse momentum 
     & -- & $\B$  & $\pt$ & $<0.1 \sqrt{s}$ \\ 
longitudinal momentum 
     & -- & $E_{\rm lost}$  & $\pl$ & $< E_1 , E_2$ \\ 
\hline
\end{tabular}
\caption{Cuts for the kinematic event selection. The cut variables
depend on the class, see text. For class \ca\ events no cuts on the 
missing longitudinal and transverse momenta are applied.}
\label{tab:kincuts}
\end{center}
\end{table}

\subsection{Neutral event selection}

Events without charged particles produced at the primary interaction point are
referred to here as neutral events. They are selected using information 
about the hits in the two main tracking chambers CJ and CV. 
A photon candidate is said to have hits in one of these chambers
if more than about 50\% of the innermost wires in the sector pointing
to the photon have hits. For CV the 12 axial layers are taken into
account and for CJ the inner 16 layers.

Events are rejected in either of the following two cases:\\
{\bf 1) double veto:}
at least two photon candidates have hits in CV. \\
{\bf 2) single veto:}
any photon candidate has hits in both CV and CJ, unless it is identified
as a conversion. 
To be identified as a conversion, information from
the SI detector is necessary and therefore this condition is
restricted to $|\ct|<0.89$.
A conversion must have exactly two tracks reconstructed in 
a three dimensional cone of half angle
20$^\circ$ around the cluster, with at least one of the tracks 
assigned to the cluster. Additionally, the two SI ladders in the 
$\phi$ direction
of the cluster must have no more than two hits. Dead ladders
are counted as hits to ensure a good background rejection. 
The two sides of each ladder are counted 
separately, leading to a maximum number of four hits.
Events are rejected if any photon with $|\ct|\geq 0.89$ has hits in CV
and CJ.

\subsection{Summary}

\begin{table}[bp]
\begin{center}
\begin{tabular}{l|r|r|rrrrrrr}
\hline
cut & \multicolumn{1}{c|}{data} & 
\multicolumn{1}{c|}{$\Sigma$MC} & 
\multicolumn{1}{c}{$\ggg$} & 
\multicolumn{1}{c}{$\epem(\gamma)$} & 
\multicolumn{1}{c}{$\rm e\gamma (e)$} & 
\multicolumn{1}{c}{$\nu\bar{\nu}\gamma(\gamma)$} &
\multicolumn{1}{c}{$\rm q\bar{q}(\gamma)$} & 
\multicolumn{1}{c}{$\tau^+\tau^-(\gamma)$} \\
\hline\hline
 preselection &
 192558 & 123751 & 5826 & 107791 & 7280 & 105 & 398 & 2352 \\\hline 
 cosmic bkg. &
 133099 & 122898 & 5823 & 107697 & 7194 & 104 & 244 & 1835 \\\hline
 kinematic cuts &
 120515 & 119674 & 5809 & 107048 & 6310 & 6.6 & 130 & 370 \\\hline
 longitudinal mom. &
 108832 & 110082 & 5520 & 103833  & 539 & 0.55 & 68 & 122  \\\hline
 double veto &
 6367 & 6152 & 5505 & 68 & 515 & 0.55 & 52 & 12\\\hline
 single veto &
 5235 &  5261 & 5258 & 0.38 & 1.55 &   0.55 &  1.10 &  0.05\\
 & & $\pm$12 & $\pm$12 & $\pm$0.19 & $\pm$0.43  & $\pm$0.09 & $\pm$0.24 &  $\pm$0.03\\\hline
\end{tabular}
\caption[]{The number of events observed in 
data after the cuts, the signal expectation and the most important 
background sources indicated by their final state are given. 
The row labelled kinematic cuts contains only the cuts on the energy sum
and the missing transverse momentum;
the cut on the missing longitudinal momentum is listed separately.
The neutral event selection is split up into the double veto and the
single veto.
For the numbers of events in the final selection after the
single veto, the statistical error is also given. All Monte Carlo 
predictions are normalised to the integrated luminosity of the data.
}
\label{tab:cutflow}
\end{center}
\end{table}

The numbers of events selected and expected after each step of the
selection are given in Table~\ref{tab:cutflow}.
The preselection is very loose and there are more
events observed after the preselection than expected from the 
listed Monte Carlo samples. This excess is mainly due to cosmic-ray events and 
four-fermion events with two observed electrons.
The difference between the number of observed and expected events 
becomes much smaller after the 
rejection of cosmic-ray background. After the kinematic cuts the sample
consists mainly of Bhabha events and the signal efficiency remains at 
almost 100\%. The cut on the longitudinal momentum
rejects events with a high-energy particle along the beam direction.
For Bhabhas these are mainly events in which one electron escapes and the
second electron and a photon are observed in the detector 
($\rm e\gamma (e)$ topology). Most of the remaining Bhabha events have 
both electrons in the detector and are easily rejected by the double 
veto. After the single veto, all background levels are almost 
negligible. Besides the requirement on the longitudinal momentum,
most of the efficiency loss comes from the single veto, since
the double veto rejects only events with two converted photons.

\section{Experimental systematic errors}
\label{sec:syserr}

The efficiency of the selection is studied with a signal Monte Carlo
simulation
\cite{ref:radcor} including full detector simulation. It is found to be
independent of the centre-of-mass energy and therefore the average efficiency 
is used to correct the data. The efficiency is about 97.8\%
in the barrel region and drops to 69\% at the edge of the selection 
at $\cte=0.93$ as can be seen in Table \ref{tab:difxs}. 
The total efficiency within $\cte < 0.93$ is 92.6\%. 
Uncertainties in the simulation give rise to the 
systematic errors as discussed below. The errors are given
in \% relative to the efficiency.

\subsection{Cut values}
To assess the stability of the cuts, the cut values are varied and 
the difference between data and Monte Carlo expectation is assigned as
a systematic error. The kinematic cuts on energy sum, missing transverse and
longitudinal momenta are changed by $\pm$10\%. The cuts on the numbers 
of hits allowed in the neutral event selection are varied by one or two
hits. The resulting systematic errors are at most 0.1\%
and in total contribute 0.17\%.

\subsection{Conversion probability}
A crucial point for this 
analysis is the correct modelling of the material in the detector 
since this material leads to photon conversions. If a photon converts
before the first active detector layer, a reliable distinction between
a primary produced electron and a photon conversion is not possible.
Events with such photons must be rejected from the analysis. Events 
with photons which convert later are selected and do not influence the 
efficiency. 

The probabilities given in this paragraph are conversion probabilities
for single photons and the given errors are absolute numbers.
For $|\ct|<0.89$ the Monte Carlo prediction for the probability for a
single photon to convert before SI is 0.7\%. A systematic error of 
0.2\% is assigned to this value since the only material in front 
of the SI detector is the beam pipe which is uniform and therefore 
well modelled. There is, however, more material in the endcap 
e.g.~from SI and CV readout electronics. In the region 
$0.89<|\ct|<0.93$ the Monte Carlo prediction for the 
conversion probability in front of CV is 6 -- 9\%. Studies, for example,
of the width of the ECAL energy distribution in Bhabha events show no 
significant disagreement in the Monte Carlo description of material in
this region, and a 1\% error is assigned to the conversion probability.
For $|\ct|>0.93$ significant effects due to unmodelled material are 
clearly visible in these studies and this is the main reason
for restricting the two highest-energy photons to $|\ct|<0.93$.
A 3\% systematic error is assigned for additional photons 
with $|\ct|>0.93$. The Monte Carlo prediction for the conversion
probability in this region is 10 -- 15\%.

Most events have two back-to-back photons. These photons hit equivalent 
kinds of material in the left and right side of the detector and consequently 
their conversion probabilities are correlated. For a two-photon event this 
leads to a systematic error on the efficiency of 0.4\% in the region 
$|\ct|<0.89$. For a typical mixture of events at all angles the average 
systematic error on the efficiency is between 0.45\% and 2.01\%, 
where the 0.45\% is taken to be correlated between $\cte$ bins.

\subsection{Angular reconstruction}
Another important source of uncertainty is the reconstruction of the 
photon angle. A systematic mis-reconstruction of the angle would lead 
to distortions in the measured angular distribution. Monte Carlo 
studies show that the angular resolution is approximately 
$0.2^\circ - 0.3^\circ$, which makes a full unfolding unnecessary.
However, some systematic shifts of up to $0.15^\circ$ are predicted 
leading to a bias of the measured angular distribution.
The Monte Carlo prediction of this bias is corrected via the 
efficiency bin-by-bin and is included in the efficiency given 
in \mbox{Table \ref{tab:difxs}}. Since the angular reconstruction for photons 
cannot be cross-checked with the data the full size of the
predicted bias is taken as a systematic error on the efficiency. 
For most of the central region this error is
negligible, at other angles the error is between 2\% and 3\%.  
For the calculation of the total cross-section, these effects are
only relevant at the edge of the phase-space of the selection,
leading to an error of 0.46\% corresponding to a 
shift of $0.15^\circ$ at $\cte=0.93$.

While the event angle is mostly affected by shifts that are symmetric on
both sides of the detector, the acollinearity is changed by shifts that
are different on the left and right side. The error arising from such 
shifts is $0.1^\circ$.

\subsection{Background}
The expected background is very small. The Monte Carlo expectation 
for Standard Model processes, summarised in Table~\ref{tab:cutflow},
is less than 4 events. Selected events in which one photon has 
hits in CV are scanned and the Standard Model background is estimated 
to be less than 10 events. Background from cosmic-ray events is estimated to be 
also less than 10 events by scanning selected events with at least one track 
segment in the muon chambers. For both these sources half of this estimate is
corrected while the other half is taken as a systematic error. 
Since the angular distribution of the background is poorly known,
to simplify the correction procedure, the shape of the signal distribution is 
assumed. Thus the background is corrected by increasing
the efficiency by 0.2\% corresponding to 10 events.
A systematic error of 0.1\% for each of the two sources is assigned.

\subsection{Other errors}

The efficiency loss due to overlaid cosmic-ray events\footnote{A cosmic-ray
particle in the detector which coincides with a signal event.}
and noise hits in the detector is studied with randomly triggered 
beam crossings. This leads to expected signal rejection rates of 0.22\%
due to cosmic-ray events and 0.30\% due to noise, corresponding to 12 and 16 
events respectively. These numbers are cross-checked by
scanning events and good agreement is found.
The resulting efficiency loss of 0.52\% is corrected for and
a systematic error of 0.05\% is assigned for each source.

The error due to Monte Carlo statistics enters only in the
determination of the efficiency which leads to a very small binomial 
error of 0.1\%. The radiative corrections are determined from
$3\cdot 10^9$ events, making the statistical error much smaller
than the theoretical uncertainty and it is therefore neglected.

The luminosity errors are given in Table \ref{tab:sample}. 
The errors change slightly from energy to energy and are dominated by
the systematic error. A total common error of 0.23\% is assigned in order 
to simplify the error treatment in the log likelihood fits.

The trigger efficiency of the ECAL is tested with Bhabha events
and is found to be 100\%. The corresponding error is negligible.

\subsection{Summary}
In general the experimental systematic errors for this analysis are 
small. The largest errors arise from the conversion probability, the
angular reconstruction and the luminosity. The overall systematic error 
for the total cross-section is 0.77\%. For the differential
cross-section the systematic error as a function of $\cte$ is given in 
Table \ref{tab:difxs}. For most of the central region the total
systematic error is 0.56\%, which is composed of 0.45\% from the
conversion probability, 0.23\% from the luminosity, 0.17\% from the cut 
values and small errors from the background and the efficiency loss. 
All systematic errors are independent of the 
centre-of-mass energy and fully correlated between energies.

\section{Results}
\label{sec:results}

\subsection{Cross-sections}

The number of observed events in each class is given in 
Table~\ref{tab:events} and compared to the number of
events expected from Monte Carlo simulation. There is a good agreement 
for collinear events \mbox{(class \ca )}, but there are fewer
acollinear 
events observed than expected. To judge whether this is an indication
for physics beyond the Standard Model one has to
take into consideration that the main Monte Carlo generator includes only 
${\cal O}(\alpha^3)$ terms. There is no Monte Carlo program available to calculate
events with four photons in which one travels along the beam direction,
so no expectation is given for class \cd\ events with three observed
photons. Events with four observed photons are calculated using 
FGAM~\cite{ref:fgam}.

\begin{table}[tp]
\setlength{\tabcolsep}{5mm}
\begin{center}
\begin{tabular}{l|ccccc}\hline
class & \ca & \cb & \cc & \multicolumn{2}{c}{\cd} \\
$\ng$ & $\geq2$ & 2 & 3 & 3 & $\geq 4$ \\ \hline\hline
observed & 4747 & 394 & 71 & 20 & 3 \\
expected & 4730 & 435 & 94 & -- & 5.3 \\
\hline
\end{tabular}
\caption{The number of observed and expected events in the different
classes. The Monte Carlo prediction for classes \ca\ -- \cc\ is from
RADCOR. The expectation for class \cd\ with four photons is calculated
using FGAM. There is no Monte Carlo program available to generate events with
three observed photons plus one escaping along the beam direction, hence
no prediction for class \cd\ events with three observed photons is given. }
\label{tab:events}
\end{center}
\end{table}

The most important discriminant for the classes is the acollinearity,
the distribution of which is shown in Figure~\ref{fig:acol}. 
The typical resolution is $0.35^\circ$. Discrepancies between data and
Monte Carlo simulation at $\acol < 1^\circ$ may occur for two reasons. One is that 
Monte Carlo events with a soft third photon below some energy cut-off 
are generated with two exactly
back-to-back photons. This affects the acollinearity distribution
at  $\acol < 1^\circ$ as can be seen in Figure \ref{fig:acol}. 
The other reason is the possible systematic shift of the reconstructed 
angle of about $0.1^\circ$. For larger values of $\acol$ differences 
are most likely 
due to missing higher order effects. The largest significance of 
a discrepancy between data and Monte Carlo simulation occurs for a cut of  
$\acol > 3^\circ$. In this case 1141 events are predicted with 1004
observed. Including a systematic error of 16 events due to the
mismodelling of the angle, an excess in the Monte Carlo prediction of 
3.6 standard deviations is observed. This discrepancy may be due 
to higher order effects, which can be large.

\begin{table}[p]
\begin{center}
\begin{tabular}{c@{$-$}c|*{8}{r}|rr|c}\hline
\multicolumn{2}{c|}{$E_{\rm nom}$} & 183 & 189 & 192 & 196 & 200 & 202 & 205 & 207 &
\multicolumn{2}{c|}{efficiency} & rad.cor. \\\cline{1-10}
\multicolumn{2}{c|}{$\cte$-bin} & \multicolumn{8}{c|}{events} & 
\multicolumn{1}{c}{$\epsilon$} & \multicolumn{1}{c|}{$\sigma_\epsilon / \epsilon$
}& $\R$ \\ 
\hline\hline
 0.00& 0.05&   8 &  33 &   3 &  10 &  15 &   8 &  17 &  23 & 0.9515& 0.0250& 1.0689\\
 0.05& 0.10&  21 &  37 &   5 &  18 &  10 &   6 &  14 &  20 & 0.9769& 0.0056& 1.0666\\
 0.10& 0.15&  15 &  34 &   4 &  14 &  15 &   4 &  13 &  15 & 0.9769& 0.0056& 1.0644\\
 0.15& 0.20&  10 &  40 &   7 &  15 &  12 &   3 &  11 &  22 & 0.9769& 0.0056& 1.0621\\
 0.20& 0.25&  13 &  40 &   2 &   9 &  13 &   5 &  14 &  23 & 0.9769& 0.0056& 1.0599\\
 0.25& 0.30&  14 &  43 &  10 &  11 &  13 &   2 &   8 &  27 & 0.9769& 0.0056& 1.0577\\
 0.30& 0.35&  20 &  47 &  10 &  14 &  20 &   7 &  29 &  26 & 0.9755& 0.0056& 1.0556\\
 0.35& 0.40&  22 &  41 &  15 &  22 &  21 &  17 &  12 &  23 & 0.9753& 0.0056& 1.0534\\
 0.40& 0.45&  21 &  38 &  12 &  19 &  11 &   9 &  18 &  18 & 0.9753& 0.0056& 1.0512\\
 0.45& 0.50&  13 &  49 &   8 &  22 &  15 &   6 &  17 &  31 & 0.9753& 0.0056& 1.0492\\
 0.50& 0.55&  21 &  61 &   7 &  23 &  24 &  10 &  22 &  23 & 0.9742& 0.0056& 1.0471\\
 0.55& 0.60&  23 &  72 &   6 &  25 &  25 &  18 &  21 &  37 & 0.9742& 0.0056& 1.0451\\
 0.60& 0.65&  25 &  64 &  14 &  38 &  24 &  19 &  21 &  48 & 0.9742& 0.0056& 1.0433\\
 0.65& 0.70&  24 &  96 &  18 &  35 &  33 &  17 &  37 &  68 & 0.9762& 0.0056& 1.0415\\
 0.70& 0.75&  36 & 104 &  18 &  42 &  34 &  23 &  50 &  75 & 0.9906& 0.0208& 1.0399\\
 0.75& 0.80&  48 & 131 &  23 &  59 &  53 &  19 &  47 &  81 & 0.9824& 0.0208& 1.0387\\
 0.80& 0.85&  66 & 176 &  33 &  71 &  72 &  35 &  71 &  89 & 0.9647& 0.0208& 1.0380\\
 0.85& 0.90&  78 & 255 &  32 & 102 &  75 &  48 &  84 & 145 & 0.8976& 0.0220& 1.0387\\
 0.90& 0.93&  60 & 170 &  31 &  67 &  69 &  25 &  60 &  97 & 0.6937& 0.0286& 1.0412\\
\hline  
\end{tabular}
\caption{Measured angular distributions. For each centre-of-mass energy 
and $\cte$
bin the number of observed events is given. The last columns give the
efficiency, its relative systematic error and the radiative correction
which has an error of 0.01. The efficiency includes the corrections
due to the background and the rejection because of noise and overlaid
cosmic-ray events. For the systematic
error on the efficiency the common contribution of 0.56\% is taken as
correlated between bins.}
\label{tab:difxs}
\end{center}
\end{table}

\begin{table}[tbp]
\begin{center}
\begin{tabular}{c|r@{$\pm$}c@{$\pm$}c|l}\hline
$\sqrt{s}$ & \multicolumn{4}{c}{Born cross-section $[\mbox{pb}]$} \\ 
$[\mbox{GeV}]$ & \multicolumn{3}{c|}{observed} & QED \\ \hline\hline
182.68 & 10.05&0.43&0.08 & 9.32 \\
188.63 &  8.79&0.23&0.07 & 8.74 \\
191.59 &  9.24&0.58&0.07 & 8.47 \\
195.53 &  8.43&0.34&0.07 & 8.13 \\
199.52 &  7.39&0.31&0.06 & 7.81 \\
201.63 &  7.88&0.47&0.06 & 7.65 \\
204.88 &  7.40&0.31&0.06 & 7.42 \\
206.56 &  6.78&0.23&0.05 & 7.29 \\
\hline
\end{tabular}
\caption{The measured and predicted total cross-sections at Born level
within the angular range of $\cte<0.93$.
The measured values are shown with their statistical and systematic
errors. The additional uncertainty on the theoretical prediction is about 1\%.}
\label{tab:totxs}
\end{center}
\end{table}

The measured differential cross-sections are plotted in 
Figure~\ref{fig:difxs} for six energy ranges. Information for all 
eight energies can be found in Table~\ref{tab:difxs}.
Integrating these distributions leads to the total cross-sections
which are shown as a function of $\sqrt{s}$ in 
Figure~\ref{fig:totxs} and are summarised in Table~\ref{tab:totxs}. 
The cross-sections are in very good agreement with the QED expectation.
On average
$\sigma_{\rm obs}/\sigma_{\rm QED} = 0.999 \pm 0.014 \pm 0.008$,
where the first error is statistical and the second is systematic.
This average has a $\rm \chi^2/dof = 12.7/7$ and includes the 
correlation of systematic errors. 
There is an additional error from theory of 0.01.

\subsection{Tests of alternative models}

Binned log likelihood fits 
for the alternative models are performed on the measured differential 
cross-sections taking systematic errors and their correlations 
into account. Where possible, the fit parameters, given in 
Table~\ref{tab:fitres}, are chosen such that the resulting probability 
distribution is approximately Gaussian. One-sided 95\% confidence level 
limits are obtained for the fit parameters by renormalising the 
likelihood function to the physically allowed region in the same way
as in \cite{ref:ich_189}. The quoted limits on the model parameters 
correspond to these limits on the fit parameters. 
For excited electrons two types of fits are performed. Limits are
determined on the coupling constant $f_\gamma/\Lambda$ of an excited 
electron to an electron and a photon for various masses $\mestar$. The
resulting limits shown in Figure~\ref{fig:flimit} do not depend on a 
special choice of $\Lambda$.
A limit on the mass $\mestar$ for a fixed coupling constant 
$f_\gamma = 1$ is also determined. In this case the scale $\Lambda$ 
is fixed to $\mestar$
and no fit parameter can be found to give a Gaussian 
probability distribution. Therefore the likelihood distribution is
shown in Figure~\ref{fig:elimit}. To make the Standard Model value at
$\mestar = \infty$ visible the likelihood is plotted as a function of
$\mestar^{-4}$. The limit of $\mestar > 245$~GeV is obtained at 
$\Delta\mbox{LogL} \equiv -\ln{\cal L} + \ln{{\cal L}_{\rm max}} = 1.92$.

All fit results are summarised in
Table~\ref{tab:fitres}. The limits on the model parameters are stronger
by 10\% to 20\% than our previously published limits~\cite{ref:ich_189}. 
Limits on the excited electron coupling $f_\gamma/\Lambda$ are much 
weaker than the limits obtained from direct searches
\cite{ref:excl_direct} but extend beyond the kinematic limit for 
excited electron production as can be seen in Figure
\ref{fig:estar_all}. As a comparison limits from excited electron
production via electron-$\gamma$ fusion determined by H1
\cite{ref:estar_H1} are also shown. Similar results are available from
ZEUS \cite{ref:estar_ZEUS}.

\begin{table}[th]
\begin{center}
\renewcommand{\arraystretch}{1.5}
\begin{tabular}{c|c|r@{ }l}\hline
Fit parameter & Fit result &
\multicolumn{2}{c}{95\%\ CL Limit [GeV]}\\  \hline\hline
 &  & $\Lambda_+ >$ &  371 \\
 \raisebox{2.2ex}[-2.2ex]{$\Lpm^{-4}$} &
 \raisebox{2.2ex}[-2.2ex]{$
 \left(-40.3{+37.2 \atop -36.5}\right)$ TeV$^{-4}$}
 & $\Lambda_- > $&  314  \\\hline
${\Lambda'}^{-6}$ & $ \left(-3.56{+2.84 \atop -2.77}\right)$ TeV$^{-6}$
& $\Lambda' > $&  800   \\ \hline
 &  & $\lambda = +1$: $M_s >$ &  805  \\
 \raisebox{2.2ex}[-2.2ex]{$\lambda/M_s^4$} &
 \raisebox{2.2ex}[-2.2ex]{ $ \left(0.926{+0.850 \atop -0.858}\right)
                                         $ TeV$^{-4} $ }
 & $\lambda = -1$: $M_s >$&  956 \\ \hline
 $\mestar (f_\gamma=1; \; \Lambda=\mestar) $ & 
 see Figure \ref{fig:elimit}   & $\mestar >$& 245 \\
 $(f_\gamma/\Lambda)^4 (\mestar=200 \mbox{ GeV})$ &
 $ \left(-215{+201 \atop -196}\right)$ TeV$^{-4}$ &
 \multicolumn{2}{c}{$f_\gamma/\Lambda <  4.11 \mbox{ TeV}^{-1}$} \\ \hline
\end{tabular}
\caption[]{Fit results and limits at 95\% confidence level obtained from 
binned log likelihood fits to the differential cross-sections. The model
parameters are defined in Section \ref{sec:theo}.}
\label{tab:fitres}
\end{center}
\end{table}

\subsection{Resonance production}

Of the 71 selected class \cc\ events, 64 have all three photons within
$|\ct| < 0.93$. These events are used to search for a photonically 
decaying resonance X which is produced in association with a photon 
($\rm\epem\to\gamma X \; ,\; X \to \gamma\gamma$). 
Each event is a 
candidate for three different masses corresponding to the pairing of 
the photons. For planar three photon events three-body kinematics 
can be used to calculate the energies $E_k$ of the photons: 
\be
E_k \propto \sin{\alpha_{ij}} \; ; \; E_1 + E_2 + E_3 = \sqrt{s} \;,
\label{eq:ggmass}
\ee
where $E_k$ is the energy of one photon and $\alpha_{ij}$ is the angle 
between the other two photons. This leads to a resolution on the
invariant mass of photon pairs of about 0.5 GeV.  
Calculating the mass from the observed cluster energies would result
instead in a resolution of about 3\%.

\begin{table}[bp]
\begin{center}
\begin{tabular}{c|rrrp{1cm}c|rrr}\hline
$\sqrt{s}\;[\mbox{GeV}]$ & \multicolumn{3}{c}{Mass $[\mbox{GeV}]$} & &
$\sqrt{s}\;[\mbox{GeV}]$ & \multicolumn{3}{c}{Mass $[\mbox{GeV}]$} \\
\hline
\hline
  181.73 &   51.7 &   84.1 &  152.6 & & 
  195.47 &  100.3 &  108.4 &  128.1 \\\cline{6-9} 
  182.41 &   47.7 &   83.5 &  155.0 & & 
  199.55 &   56.2 &  126.9 &  143.3 \\ 
  182.72 &   55.7 &   73.8 &  157.6 & & 
  199.56 &   43.1 &  114.3 &  157.8 \\ 
  182.75 &   58.7 &   49.5 &  165.8 & & 
  199.57 &   81.3 &   48.3 &  175.7 \\ 
  182.75 &   42.7 &   68.4 &  164.0 & & 
  195.59 &   56.0 &  106.5 &  154.2 \\ 
  182.86 &   60.1 &  118.2 &  125.9 & & 
  195.59 &   80.3 &   82.9 &  158.0 \\ 
  182.89 &   89.0 &  104.2 &  121.1 & & 
  199.41 &   90.0 &   96.8 &  149.3 \\ 
  182.70 &   51.3 &  103.2 &  141.8 & & 
  199.45 &   78.3 &  121.1 &  137.8 \\ 
  182.70 &   67.6 &  116.3 &  123.6 & & 
  199.52 &   99.3 &  116.3 &  128.1 \\\cline{6-9} 
  182.70 &   69.5 &   84.1 &  146.5 & & 
  201.58 &   40.8 &  127.3 &  150.9 \\\cline{1-4}
  188.58 &   60.4 &   50.7 &  171.3 & & 
  201.63 &   81.8 &  118.9 &  140.8 \\\cline{6-9}  
  188.59 &   43.5 &   88.8 &  160.6 & & 
  202.55 &   49.3 &   99.4 &  169.5 \\ 
  188.59 &   79.9 &   69.0 &  156.2 & & 
  203.66 &   83.1 &   78.0 &  168.8 \\ 
  188.59 &  102.2 &   78.8 &  137.5 & & 
  204.65 &   26.2 &  127.3 &  158.1 \\ 
  188.60 &   33.6 &   94.8 &  159.5 & & 
  204.73 &   51.4 &   86.8 &  178.2 \\ 
  188.61 &   76.3 &  102.7 &  138.5 & & 
  205.04 &   54.9 &   64.9 &  186.6 \\ 
  188.62 &   49.5 &   73.7 &  166.4 & & 
  205.11 &   60.5 &   78.2 &  179.7 \\ 
  188.63 &   66.5 &   45.9 &  170.4 & & 
  205.13 &   65.0 &   84.6 &  175.2 \\ 
  188.63 &   51.1 &   76.0 &  164.9 & & 
  205.14 &   62.5 &   93.9 &  171.4 \\ 
  188.64 &  105.7 &   97.8 &  121.8 & & 
  205.35 &   37.7 &  137.0 &  148.3 \\\cline{6-9}  
  188.65 &   37.9 &   98.7 &  156.3 & & 
  205.81 &   73.1 &  120.4 &  150.0 \\ 
  188.66 &  115.3 &   99.0 &  111.8 & & 
  206.03 &   62.4 &   84.4 &  177.3 \\ 
  188.66 &   35.7 &   94.4 &  159.4 & & 
  206.17 &   63.1 &   64.2 &  185.5 \\ 
  188.66 &   67.6 &   65.8 &  163.4 & & 
  206.17 &   58.4 &  103.1 &  168.7 \\ 
  188.67 &   93.0 &   77.9 &  144.5 & & 
  206.39 &   54.2 &   92.4 &  176.4 \\ 
  188.67 &   35.1 &   96.4 &  158.3 & & 
  206.55 &   66.0 &  116.7 &  157.2 \\ 
  188.67 &   54.7 &   86.3 &  158.6 & & 
  206.56 &   93.8 &  104.1 &  151.8 \\ 
  188.67 &   65.4 &   92.8 &  150.7 & & 
  206.57 &   83.3 &   65.0 &  177.5 \\ 
  189.01 &   34.4 &   99.3 &  157.1 & & 
  206.58 &  105.5 &  125.5 &  125.7 \\\cline{1-4} 
  191.52 &   41.2 &   84.4 &  166.9 & & 
  206.61 &   69.6 &  108.7 &  161.4 \\ 
  191.63 &   43.2 &  116.9 &  145.6 & & 
  207.96 &   69.9 &   84.9 &  176.5 \\\cline{1-4} 
  195.47 &   61.5 &   83.3 &  165.8 & &
  208.03 &   43.1 &  112.1 &  169.9 \\ 
\hline
\end{tabular}
\caption{The three invariant masses for planar three photon events.
The events are grouped according to the energy ranges given in 
Table~\ref{tab:sample}.}
\label{tab:mass}
\end{center}
\end{table}

For the signal it is assumed that the resonance X has isotropic
production and decay distributions. The QED
background on the other hand is peaked in the forward direction. 
Since the QED Monte Carlo generator appears to overestimate the production of 
three-photon events the background is scaled to the number
of observed events.

The efficiency for resonance production can be separated into two parts.
The first part accounts for the restricted phase-space of the selection
summarised in Equation \ref{eq:gggspace}:
\be
|\ct_i|<0.93,\; i=1,2,3 \; ; \; \acol > 10^\circ \; .
\label{eq:gggspace}
\ee
This efficiency depends strongly on 
the mass, $M_{\rm X}$, of the resonance. At a centre-of-mass energy 
of 207 GeV it is
between 64\% and 72\% for masses in the range 70 GeV to 170 GeV. 
The mass range is limited by the acollinearity cut.
The second part is the reconstruction efficiency for $\rm X\gamma$ events 
within the phase-space defined by Equation~\ref{eq:gggspace}; this 
depends only weakly on $M_{\rm X}$ and the centre-of-mass energy. 
It is 96.3\% which is larger than the efficiency
for QED events because of the different angular distribution.

The observed distribution of the invariant mass of photon pairs is shown
in Figure~\ref{fig:masslim} together with the distribution expected for
a signal at $M_{\rm X} = 131.5$ GeV. 
A list of the three invariant masses per event can be
found in Table~\ref{tab:mass}. Limits at 95\% confidence level
on the product of $\rm X\gamma$ production
cross-section and branching ratio are derived using a method
of fractional event counting~\cite{ref:bock},
where the weight of an event is determined
according to the resolution and the difference between hypothetical and
reconstructed masses. The intrinsic width of the resonance X is assumed to be
negligible. The limits are calculated assuming that the cross-section 
is independent of $\sqrt{s}$. 
If the production cross-section depends on $\sqrt{s}$, the limit
can be used with respect to the luminosity weighted centre-of mass 
energy of $\langle\sqrt{s}\rangle = $196.6~GeV. For most production 
cross-sections (e.g. $\sigma\propto 1/s$ or $\propto \ln{s}$) the 
difference from the correctly scaled limit is less than 0.8\% for masses
in the range 40~GeV to 170~GeV. For a threshold behaviour like 
$\sigma\propto(1-\frac{M_{\rm X}^2}{s})^3$ \cite{ref:ram} this holds only up to 
$M_{\rm X} <$~145 GeV. The limits are about a factor two better than our 
previously published results~\cite{ref:ich_189}.

\section{Conclusion}
\label{sec:concl}

Total and differential cross-sections for the process $\eeggg$ have 
been measured at high energies 
with a large data set. The data are in good agreement with the QED
expectation, though some discrepancies in the acollinearity distribution
are observed, which are attributed to
missing higher order effects in the $\O(\alpha^3)$ model
calculations. Strong constraints on models predicting deviations from 
QED are obtained which are summarised in Table~\ref{tab:fitres}. 
Lower limits at 95\% confidence level are placed on cut-off parameters
$\Lpm$ of about 340 GeV and on the scale of gravity in extra dimensions
of about 880 GeV. Excited electrons must be heavier than 245 GeV if 
the relative strength of the $\rm e^\ast e\gamma$ vertex is 
$f_\gamma/\Lambda = 1/\mestar$.
Limits on this coupling are placed for a range of excited electron
masses. In the mass spectrum of photon pairs no indication for a 
narrow resonance X
is found leading to limits on cross-section times branching ratio
for $\rm X\gamma$ production with $\rm X\to\gamma\gamma$ of about 
0.02 pb assuming isotropic production and decay distributions.

\section*{Acknowledgements:}

We particularly wish to thank the SL Division for the efficient operation
of the LEP accelerator at all energies
 and for their close cooperation with
our experimental group.  In addition to the support staff at our own
institutions we are pleased to acknowledge the  \\
Department of Energy, USA, \\
National Science Foundation, USA, \\
Particle Physics and Astronomy Research Council, UK, \\
Natural Sciences and Engineering Research Council, Canada, \\
Israel Science Foundation, administered by the Israel
Academy of Science and Humanities, \\
Benoziyo Center for High Energy Physics,\\
Japanese Ministry of Education, Culture, Sports, Science and
Technology (MEXT) and a grant under the MEXT International
Science Research Program,\\
Japanese Society for the Promotion of Science (JSPS),\\
German Israeli Bi-national Science Foundation (GIF), \\
Bundesministerium f\"ur Bildung und Forschung, Germany, \\
National Research Council of Canada, \\
Hungarian Foundation for Scientific Research, OTKA T-029328, 
and T-038240,\\
Fund for Scientific Research, Flanders, F.W.O.-Vlaanderen, Belgium.\\

\bibliographystyle{Lep2Rep}
\bibliography{pr363}


\begin{figure}[p]
   \vspace*{-1cm}
   \begin{center} \mbox{
          \epsfxsize=16.0cm
           \epsffile{pr363_01.eps}
           } \end{center}
           \vspace*{-1cm}
\caption[ ]{Last OPAL $\eeggg$ event, taken three minutes before the
final shut-down of LEP. Two high-energy clusters are detected in the 
electromagnetic calorimeter.  This is a class \cb\ event with 
acollinearity $\acol=17^\circ$. One photon has converted between 
CV and CJ; the two corresponding tracks are visible in the tracking chambers.
}
\label{fig:event}
\end{figure}

\begin{figure}[p]
   \vspace*{-1cm}
   \begin{center} \mbox{
          \epsfxsize=16.0cm
           \epsffile{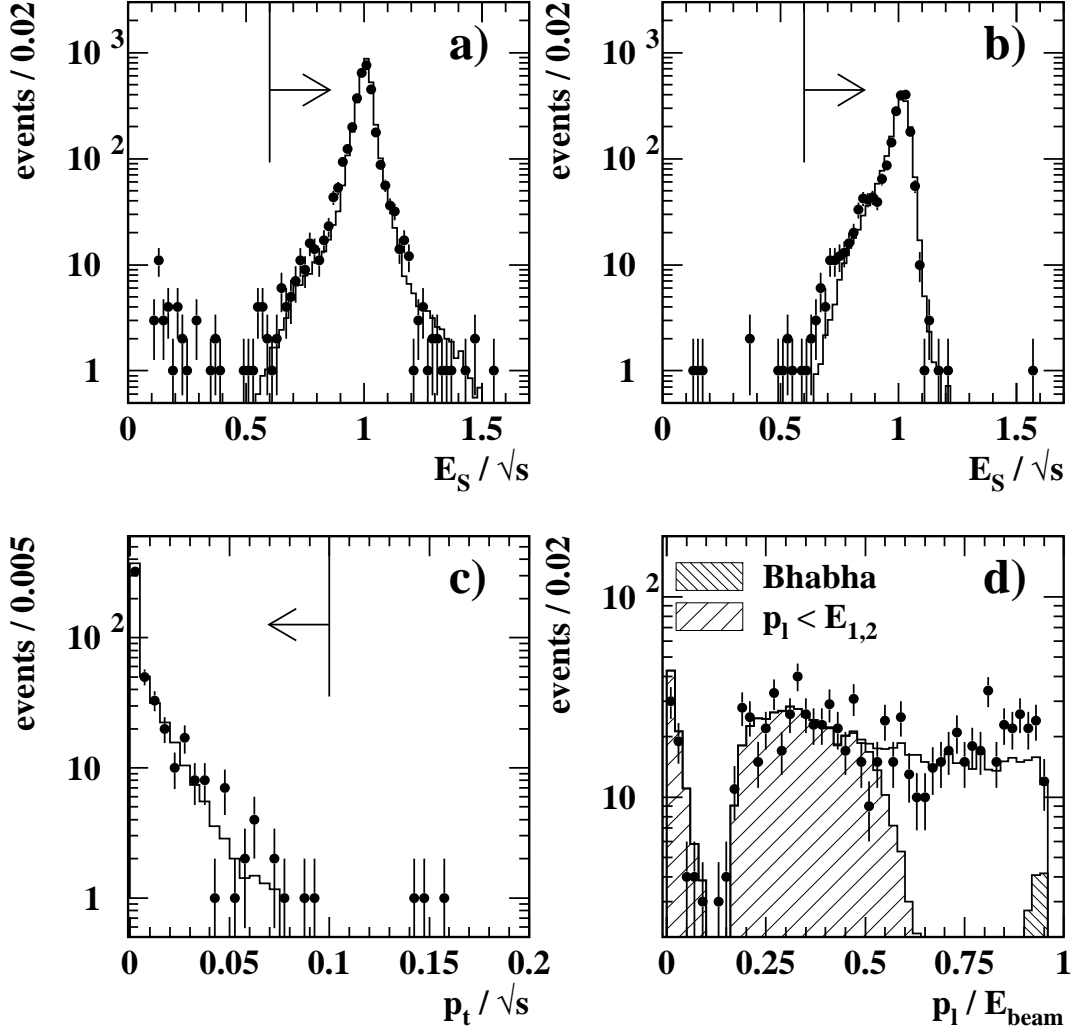}
           } \end{center}
           \vspace*{-1cm}
\caption[ ]{Distributions used in the kinematic event selection.
The energy sum for events of all classes is shown for barrel and endcap 
regions separately in plots a) $\cte < 0.81$ and b) $\cte\geq 0.81$. 
For events that are not in class \ca\ plot c) shows the transverse 
momentum and d) the longitudinal momentum. The points are the OPAL data 
after application of all cuts except that on the quantity which is 
plotted. The histograms show the Monte Carlo
expectation. The arrows in a), b) and c) show the positions of the cuts.
For the longitudinal momentum there is no cut at a specific value,
events are selected if $\pl < E_{1,2}$ as indicated by the Monte Carlo
distribution shown as the shaded histogram. Background comes mainly 
from cosmic-ray events. In plot d) some Bhabha background is visible.
}
\label{fig:kin}
\end{figure}

\begin{figure}[p]
   \vspace*{-1cm}
   \begin{center} \mbox{
          \epsfxsize=16.0cm
           \epsffile{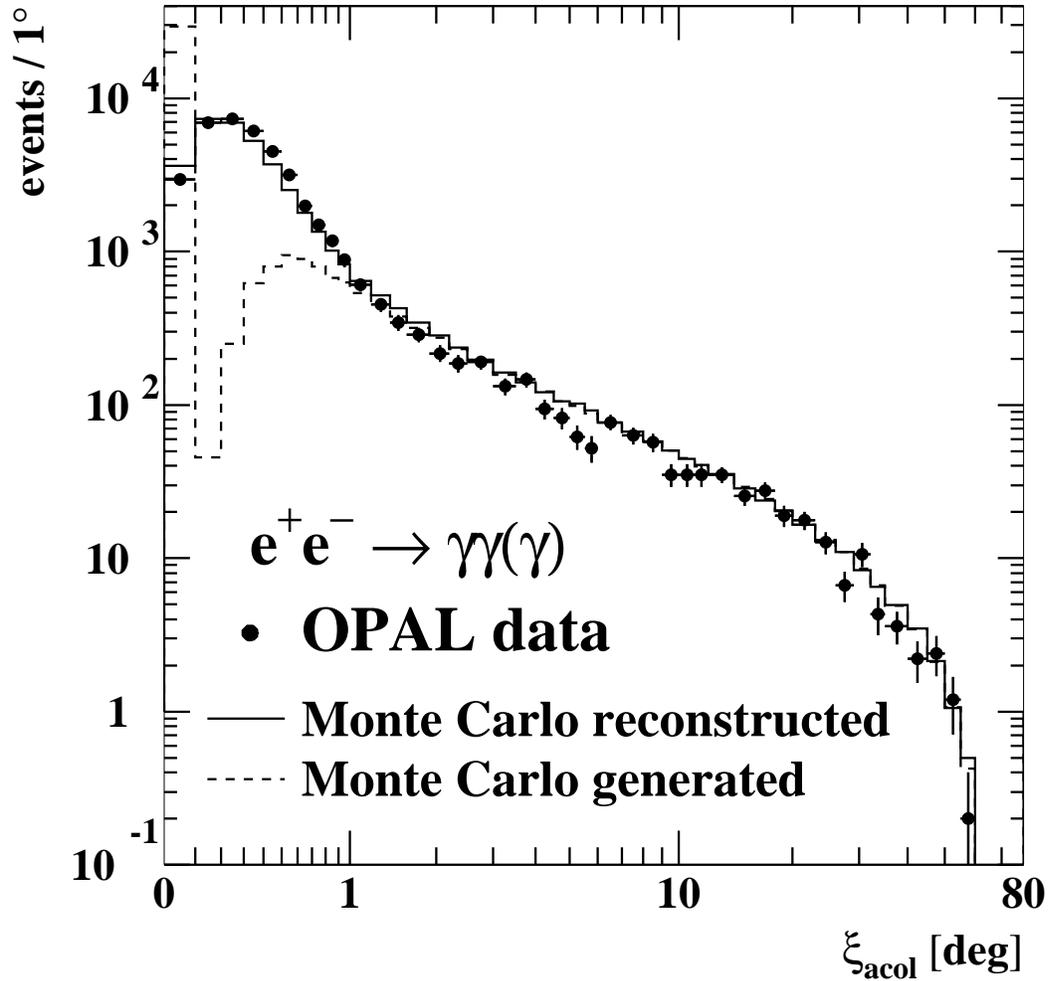}
           } \end{center}
           \vspace*{-1cm}
\caption[ ]{Acollinearity $\acol$ for all selected events. 
Because of the non-uniform bin size the entries are normalised to
events/1$^\circ$.
The measured distribution is compared to the $\O(\alpha^3)$ Monte Carlo
prediction including full detector simulation. The Monte Carlo distribution 
at generator level is also shown. The resolution is 
about $0.35^\circ$ and the possible systematic bias on the angle is 
around 0.1$^\circ$.
}
\label{fig:acol}
\end{figure}

\begin{figure}[p]
   \vspace*{-1cm}
   \begin{center}
      \mbox{
          \epsfxsize=16.0cm
          \epsffile{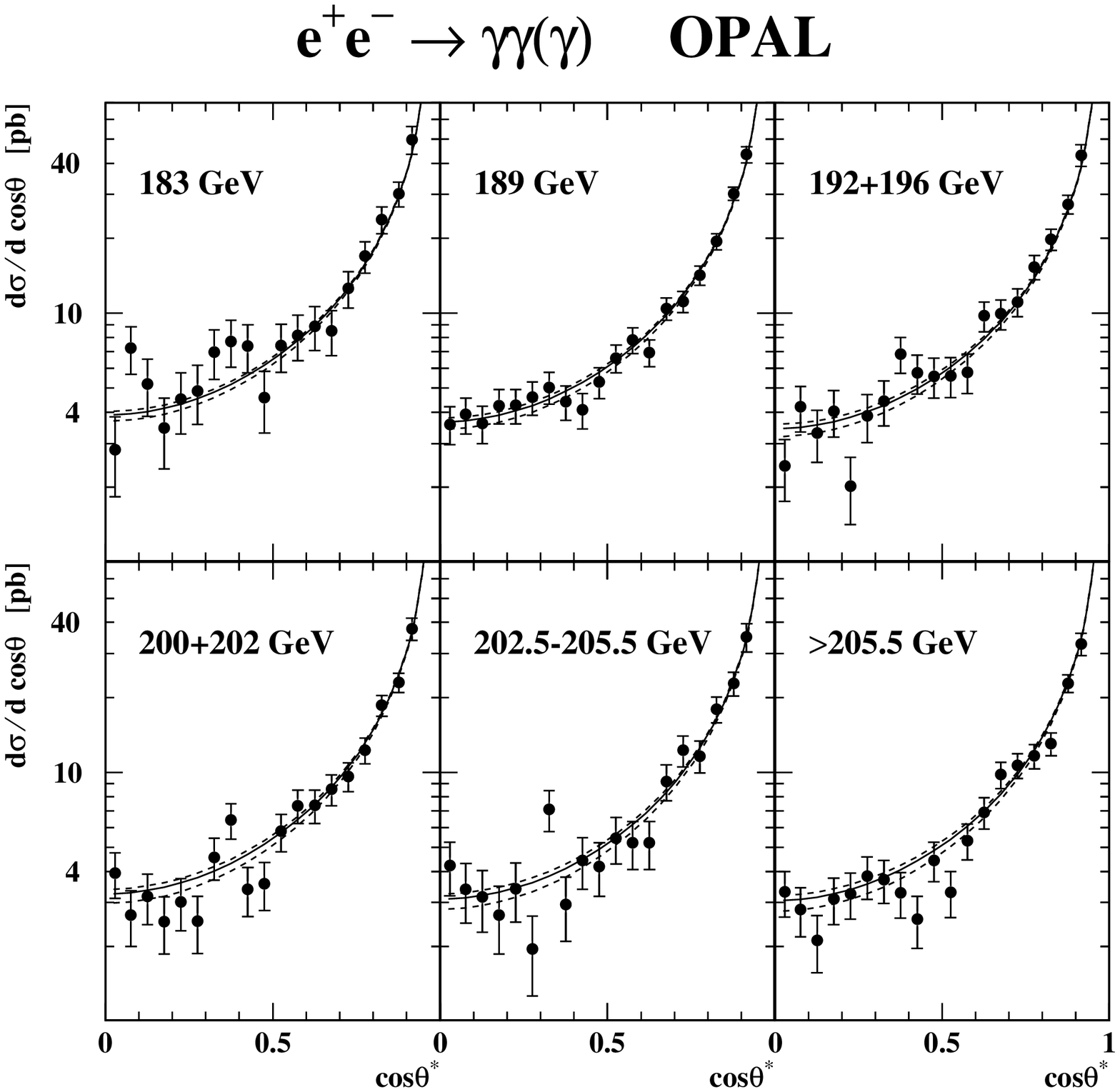}
           }
   \end{center}
\caption[ ]{The measured differential cross-section at the Born level 
for the process 
$\epem\to\g\g(\g)$ for six ranges of centre-of-mass energy.
The points show the number of observed events corrected for 
efficiency and radiative effects. The solid curve corresponds to
the Born-level QED prediction. The dashed lines represent the 95\% 
confidence level interval from the fit to the function given in 
Equation \ref{lambda}.  }
\label{fig:difxs}
\end{figure}

\begin{figure}[p]
   \begin{center} \mbox{
          \epsfxsize=15.0cm
           \epsffile{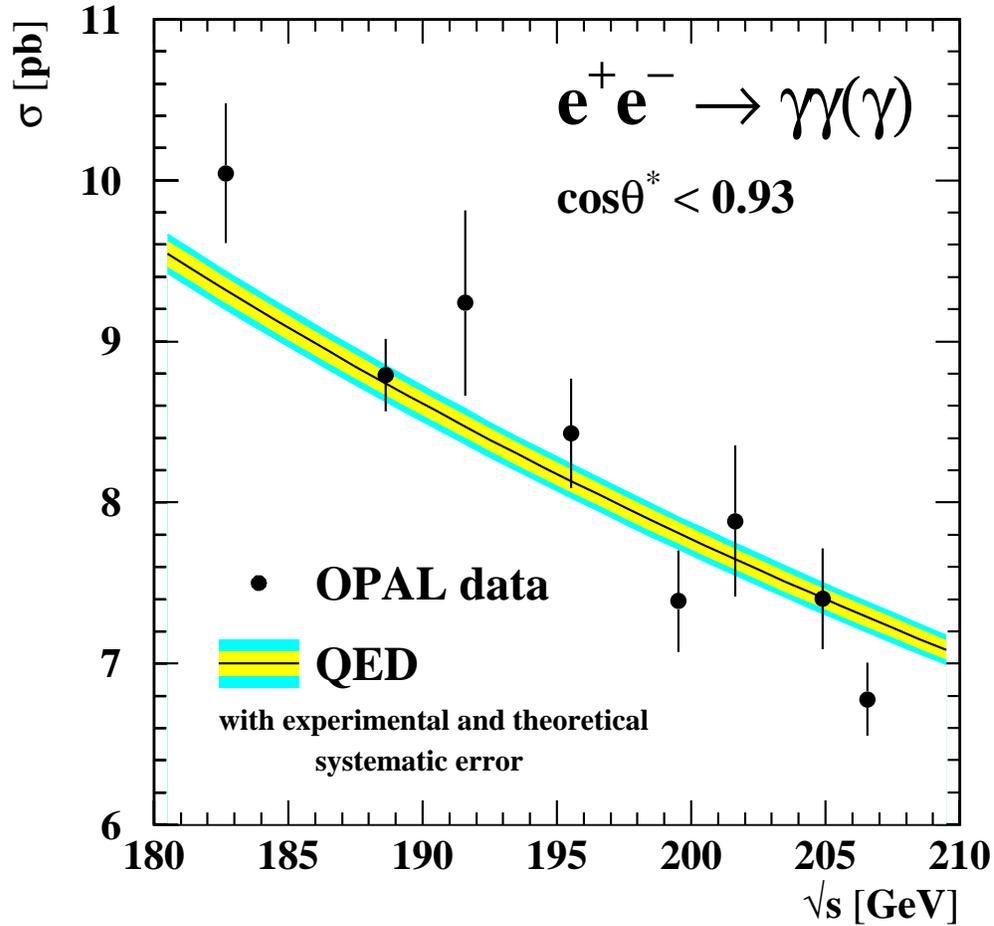}
           } \end{center}
\caption[ ]{Total cross-section at the Born level for the process $\eeggg$ 
with $\cte < 0.93$. The curve corresponds to the Born-level QED expectation. 
The data are corrected for efficiency loss and higher-order effects. 
The errors shown for the measurements are statistical only. The
systematic errors are correlated between energies and are plotted as a 
band around the QED expectation. The inner band represents the 
experimental error and the outer band includes the theoretical error.
}
\label{fig:totxs}
\end{figure}

\begin{figure}[p]
   \begin{center}
\vspace*{-2cm}
      \mbox{
          \epsfxsize=16.0cm
          \epsffile{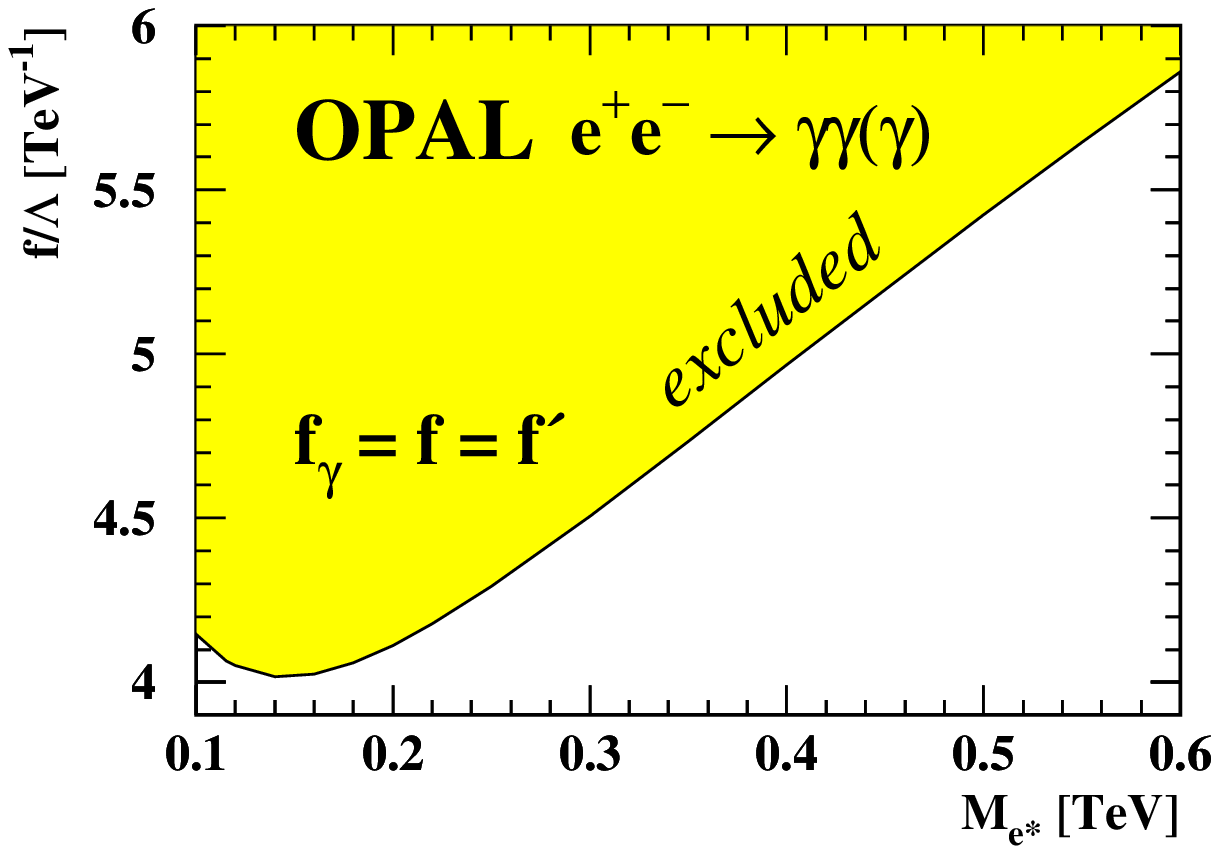}
           }
   \end{center}
           \vspace*{-1.5cm}
\caption[ ]{Upper limit at 95\%\ confidence level on the coupling $f_\gamma/\Lambda$ as a 
function of the mass $M_{\rm e^{\ast}}$ of an excited electron.
}
\label{fig:flimit}
%
   \begin{center}
\vspace*{-1cm}
      \mbox{
          \epsfxsize=16.0cm
          \epsffile{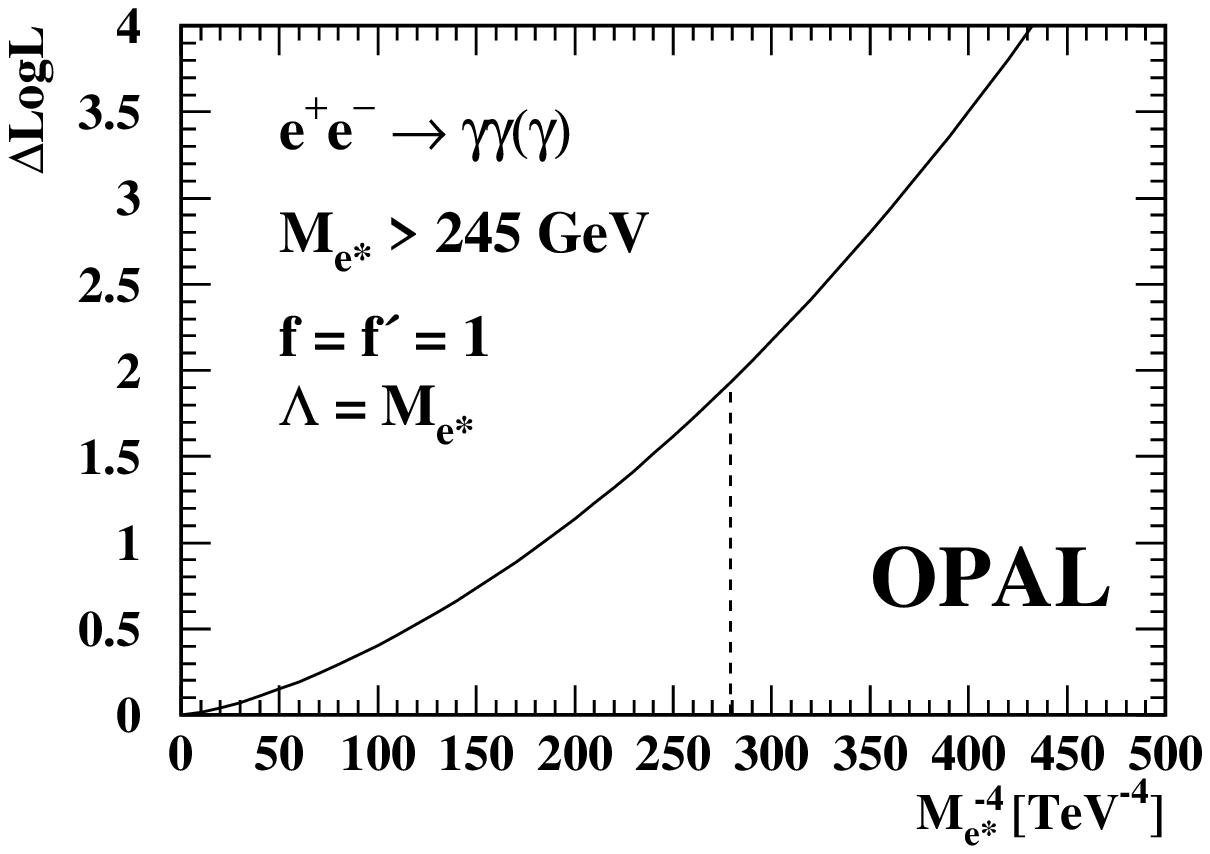}
           }
   \end{center}
           \vspace*{-1.5cm}
\caption[ ]{Likelihood difference as a function of
$\mestar^{-4}$ for fixed $f_\gamma = 1$ and $\Lambda=\mestar$. 
The limit obtained at 
$\Delta\mbox{LogL} = 1.92$ is $\mestar > 245$ GeV.
}
\label{fig:elimit}
\end{figure}

\begin{figure}[p]
   \vspace*{-2cm}
   \begin{center} \mbox{
          \epsfxsize=16.0cm
           \epsffile{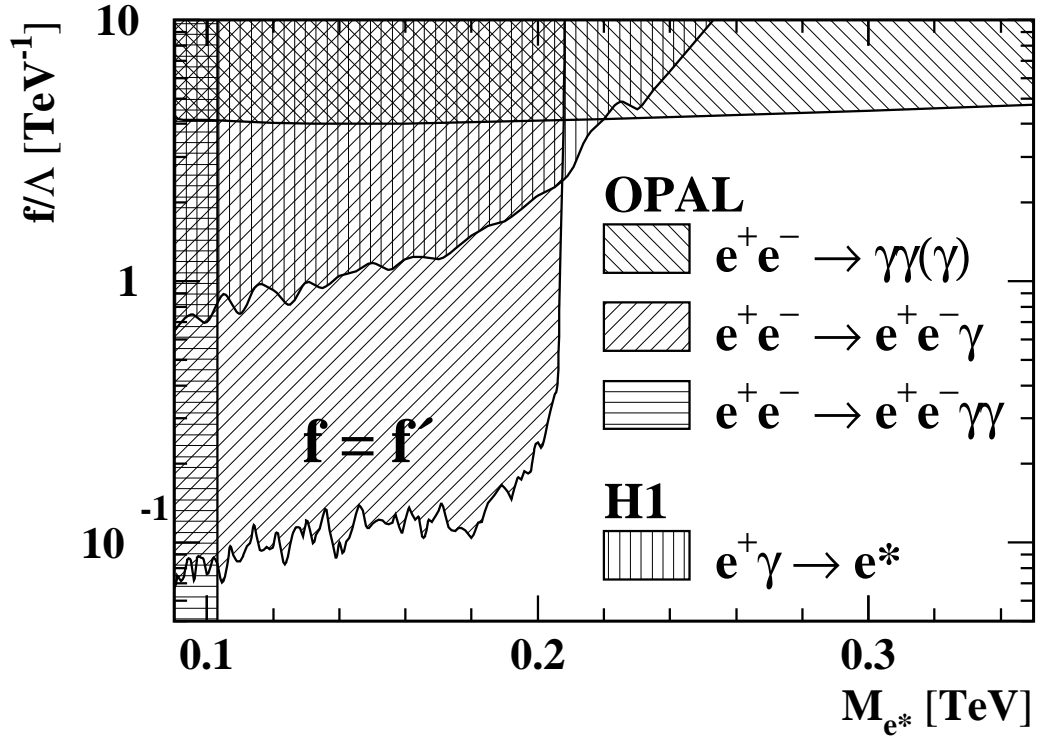}
           } \end{center}
\caption[ ]{Comparison of limits on the coupling of excited electrons.
OPAL results are given from this analysis and the two channels of the
direct search \cite{ref:excl_direct} together with the results from 
H1 \cite{ref:estar_H1}. The shaded regions are excluded at 95\%
confidence level in each case.}
\label{fig:estar_all}
\end{figure}

\begin{figure}[p]
   \vspace*{-2cm}
   \begin{center} \mbox{
          \epsfysize=20.0cm
           \epsffile{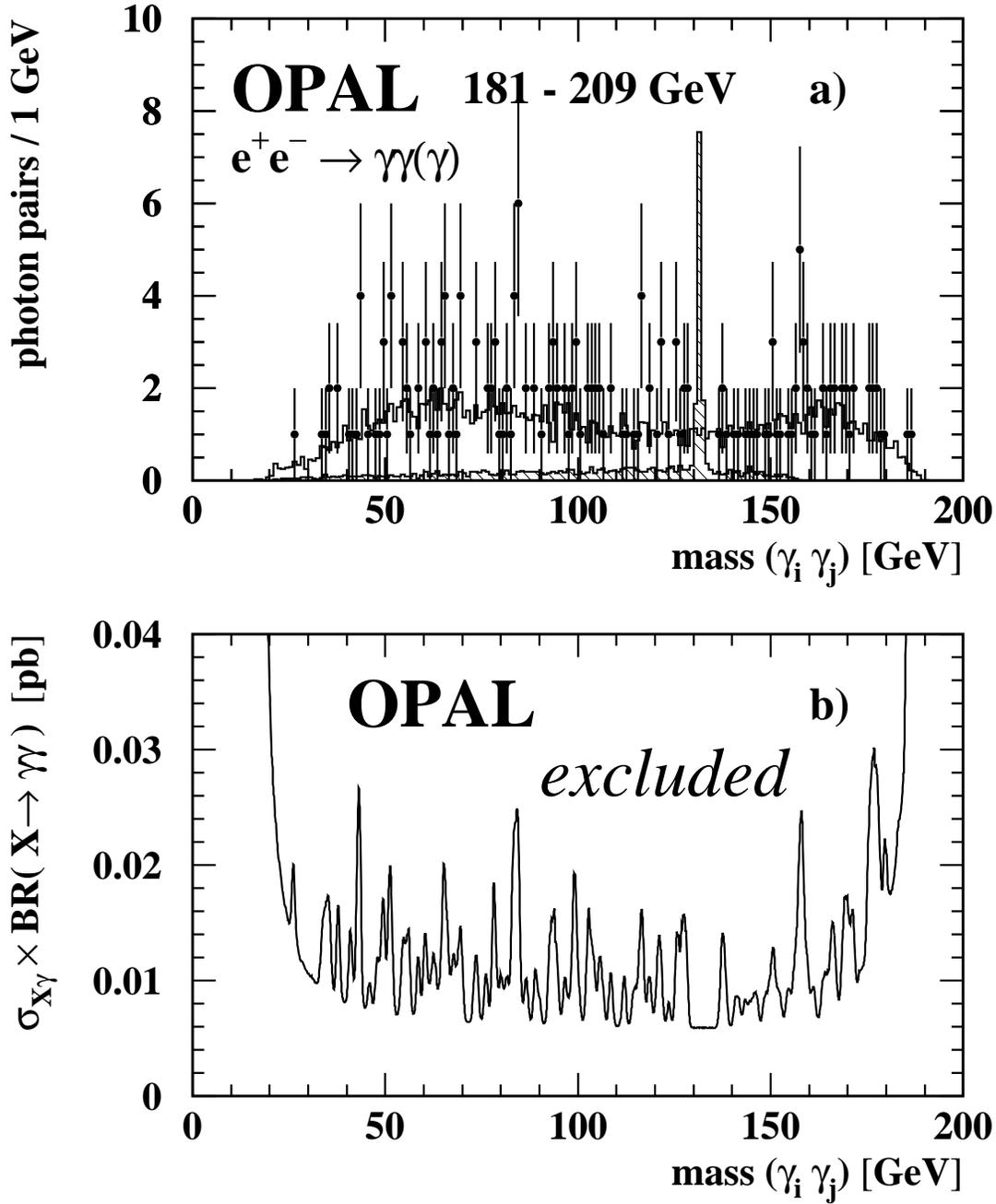}
           } \end{center}
\caption[ ]{Results of a search for resonance production in class 
\cc\ events.
a) shows the invariant mass of photon pairs for data (points) and 
for the $\eegg\gamma$ Monte Carlo expectation (histogram) scaled to 
the number of observed events. There are three photon pairs per event.
A hypothetical signal with $\sigma_{\rm X\gamma} \times \mbox{BR} = 0.02$ pb
and a mass $M_{\rm X}=131.5$ GeV is superimposed (hatched histogram).
The binning is chosen to match the expected mass resolution.
b) shows the upper limit at 95\%\ confidence level for the product of production
cross-section and 
branching ratio for the process $\rm \epem\to X \g$, $\rm X\to\g\g$ 
as a function of the mass of the resonance X. }
\label{fig:masslim}
\end{figure}

\end{document}

%% file: collaboration.tex
\begin{center}{\Large        The OPAL Collaboration
}\end{center}\bigskip
\begin{center}{
G.\thinspace Abbiendi$^{  2}$,
C.\thinspace Ainsley$^{  5}$,
P.F.\thinspace {\AA}kesson$^{  3}$,
G.\thinspace Alexander$^{ 22}$,
J.\thinspace Allison$^{ 16}$,
P.\thinspace Amaral$^{  9}$, 
G.\thinspace Anagnostou$^{  1}$,
K.J.\thinspace Anderson$^{  9}$,
S.\thinspace Arcelli$^{  2}$,
S.\thinspace Asai$^{ 23}$
% ...
}\end{center}\bigskip
\bigskip
$^{  1}$School of Physics and Astronomy, University of Birmingham,
Birmingham B15 2TT, UK
\newline
$^{  2}$Dipartimento di Fisica dell' Universit\`a di Bologna and INFN,
I-40126 Bologna, Italy
\newline
$^{  3}$Physikalisches Institut, Universit\"at Bonn,
D-53115 Bonn, Germany
\newline
% ...
$^{  5}$Cavendish Laboratory, Cambridge CB3 0HE, UK
\newline
% ...
$^{  9}$Enrico Fermi Institute and Department of Physics,
University of Chicago, Chicago IL 60637, USA
\newline
% ...
$^{ 16}$Department of Physics, Schuster Laboratory, The University,
Manchester M13 9PL, UK
\newline
% ...
$^{ 22}$Department of Physics and Astronomy, Tel Aviv University,
Tel Aviv 69978, Israel
\newline
$^{ 23}$International Centre for Elementary Particle Physics and
Department of Physics, University of Tokyo, Tokyo 113-0033, and
Kobe University, Kobe 657-8501, Japan
\newline
% ...
\newpage